\newif\ifarxiv\arxivtrue%

\ifarxiv%
\pdfoutput=1
\fi

\documentclass[runningheads,orivec]{llncs}

\usepackage{etoolbox}
\apptocmd{\sloppy}{\hbadness 10000\relax}{}{} 

\RequirePackage[paperheight=235mm,paperwidth=155mm,textwidth=12.2cm,textheight=19.3cm,inner=55pt]{geometry}

\ifarxiv%
\makeatletter
\def\@citecolor{blue}%
\def\@urlcolor{blue}%
\def\@linkcolor{blue}%

\def\orcidID#1{\href{http://orcid.org/#1}{\protect\raisebox{-1.25pt}{\protect\includegraphics{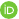}}}}
\makeatother
\fi%

\usepackage{amsmath}
\usepackage{amssymb}
\usepackage{stmaryrd}
\usepackage{mathpartir}

\usepackage{silence}
\usepackage{hyperref}
\usepackage[T1]{fontenc}

\usepackage[nameinlink]{cleveref}

\usepackage{algorithm2e}
\SetKwSty{texttt}

\newcommand{\BExp}{\mathsf{BExp}}
\newcommand{\Exp}{\mathsf{Exp}}
\newcommand{\GExp}{\mathsf{GExp}}
\newcommand{\At}{\mathsf{At}}
\newcommand{\sem}[1]{\llbracket#1\rrbracket}
\newcommand{\id}{\mathsf{id}}

\usepackage{subcaption}

\usepackage{tikz}
\usetikzlibrary{arrows}
\usetikzlibrary{automata}
\usetikzlibrary{positioning}

\begin{document}
\title{On Propositional Program Equivalence \texorpdfstring{\\}{} (extended abstract)}

\author{Tobias Kappé\inst{1}\orcidID{0000-0002-6068-880X}} 
\authorrunning{T.~Kappé}

\institute{LIACS, Leiden University \email{t.w.j.kappe@liacs.leidenuniv.nl}}

\maketitle

\begin{abstract}
General program equivalence is undecidable.
However, if we abstract away the semantics of statements, then this problem becomes not just decidable, but practically feasible.
For instance, a program of the form \texttt{if $b$ then $e$ else $f$} should be equivalent to \texttt{if not $b$ then $f$ else $e$} --- no matter what $b$, $e$ and $f$ are.
This kind of equivalence is known as \emph{propositional equivalence}.
In this extended abstract, we discuss recent developments in propositional program equivalence from the perspective of (Guarded) Kleene Algebra with Tests, or (G)KAT\@. 
\end{abstract}

\section{Introduction}

Modern programming languages offer a wealth of techniques to implement a desired computation.
The resulting flexibility also means that some changes to a program do not (and should not) change its semantics.
As such, there are circumstances where one would like to check whether a rearrangement of the source code is equivalent to the original.
Because this problem is undecidable in general, there are two options: we can try to focus on specific fragments of programming languages where equivalence is decidable, or we can adjust our notion of equivalence to make it decidable.
This extended abstract is about the latter approach, in particular using \emph{(Guarded) Kleene Algebra with Tests}~\cite{kozen-1997,kozen-tseng-2008,smolka-etal-2020}.

More specifically, the notion of equivalence under study is called \emph{propositional equivalence}.
In a nutshell, two programs are propositionally equivalent if their semantics coincide, independently of the interpretation of the primitive statements.
As an example, consider a program in an imperative programming language with traditional branches and loops, such as \texttt{if $b$ then \{ $e$; while $b$ do $e$ \}}.
It should be clear that this program has the same effect as \texttt{while $b$ do $e$}, regardless of the actual test filled in for $b$, or the actual statement (or indeed, program) that appears at $e$.\footnote{For the sake of simplicity, let's assume that $e$ does not contain any context-dependent control flow, like \texttt{break}. This assumption can be relaxed~\cite{zhang-etal-2025}.}
Of course, equivalences like this one are relatively easy to grasp, and most programmers will have an intuitive understanding of them.
On the other hand, equivalences between more involved programs are harder to justify by themselves; in such cases, it would be preferable if we could establish them as consequences of simpler rules, or check them mechanically.

As an example of such a nontrivial equivalence, consider a program meant to visit the nodes of a binary tree in-order, i.e., by visiting each node after its left child, but before its right child.
A plausible implementation appears in \Cref{algorithm:triple-loop}.

\begin{figure}[t]
\centering
\begin{subfigure}[b]{0.45\textwidth}
\begin{algorithm}[H]
\texttt{s := new stack()}\;
\texttt{node := root}\;
\While{\textnormal{\texttt{node != nil}}}{%
    \texttt{s.push(node)}\;
    \texttt{node := node.left}\;
}
\While{\textnormal{\texttt{!s.empty}}}{%
    \texttt{node = s.pop()}\;
    \texttt{visit(node)}\;
    \texttt{node = node.right}\;
    \While{\textnormal{\texttt{node != nil}}}{%
        \texttt{s.push(node)}\;
        \texttt{node := node.left}\;
    }
}
\end{algorithm}
\caption{Triple-loop version}\label{algorithm:triple-loop}
\end{subfigure}
\begin{subfigure}[b]{0.51\textwidth}
\begin{algorithm}[H]
\texttt{s := new stack()}\;
\texttt{node := root}\;
\While{\textnormal{\texttt{node != nil or !s.empty}}}{%
    \eIf{\textnormal{\texttt{node != nil}}}{%
        \texttt{s.push(node)}\;
        \texttt{node = node.left}\;
    }{%
        \texttt{node = s.pop()}\;
        \texttt{visit(node)}\;
        \texttt{node = node.right}\;
    }
}
\end{algorithm}
\caption{Single-loop version}\label{algorithm:single-loop}
\end{subfigure}
\caption{Two algorithms for an in-order walk of a binary tree.}\label{figure:algorithms}
\end{figure}

While this implementation is not entirely straightforward, it implements a correct in-order traversal.
Nevertheless, there is something to be desired: the first loop is the same as the third (inner) loop.
We could abstract this code into a function, but a different approach is possible too, unifying the three loops into one.
The underlying idea is that each ``step'' of the algorithm either visits the current node, or remembers a node to be visited later, with the latter action taking priority.
This yields the implementation in \Cref{algorithm:single-loop}~\cite[\S2.3.1]{knuth-1997}.

If these implementations are equivalent, then the second program must also be correct --- but how do we prove this?
Perhaps surprisingly, propositional equivalence can help: regardless of what \texttt{s.push(node)} and other actions do, these programs achieve the same outcome. 
Put differently: \texttt{$e$; while $b$ do $f$; while $c$ do \{ $g$; while $b$ do $f$ \}} is equivalent to \texttt{$e$; while $b$ or $c$ do \{ if $b$ then $f$ else $g$ \}}, for all programs $e$, $f$ and $g$, and tests $b$ and $c$.
In fact, these programs perform the same actions (stack manipulations, node visits, variable assignments), in the same order, when given the same input.
As hinted at, we would like to be able to prove this far-from-trivial equivalence using simpler laws.
This is where \emph{Guarded Kleene Algebra with Tests}~\cite{kozen-tseng-2008,smolka-etal-2020}, or \emph{GKAT}, can help, as an algebraic system to prove equivalences such as this one.

The extended abstract will go over the syntax, semantics, and metatheory of GKAT\@.
The objective is not to give a comprehensive, textbook-like treatment, but rather to serve as a starting point for further study to those interested in learning more, with plenty of references for further reading.
We start in \Cref{section:kat} with an overview of Kleene Algebra with Tests; next, we delve into the specifics of GKAT in \Cref{section:gkat}.
We discuss connections with process algebra in \Cref{section:pa}, and conclude in \Cref{section:horizons} with some suggestions for future work.

\section{Kleene Algebra with Tests}\label{section:kat}

To properly explain GKAT, we need to spend some time discussing its predecessor, \emph{Kleene Algebra with Tests} (\emph{KAT})~\cite{kozen-1996,kozen-1997,kozen-smith-1996,cohen-kozen-smith-1996}.
As the name might suggest, KAT is an extension of an earlier system known as Kleene Algebra~\cite{kozen-1994,conway-1971,krob-1990,salomaa-1966}; it is a powerful algebraic system to reason about propositional equivalence between programs that include non-deterministic composition and loops.
This non-determinism makes KAT extremely expressive.
However, as we will see, it also comes at the cost of a high (theoretical) complexity when deciding equivalence~\cite{cohen-kozen-smith-1996} (although a practically feasible equivalence checking algorithm does exist~\cite{pous-2015}).

The syntax of KAT is defined in two stages.
First, there are the \emph{tests}, which represent a propositional abstraction of the conditions that might occur in branches (such as \texttt{node != nil} in our motivating example). 
These are generated by a set of \emph{primitive tests} $T$, which we fix for the remainder, as follows:
\[
    \BExp \ni b, c ::= 0 \mid 1 \mid t \in T \mid b + c \mid b \cdot c \mid \overline{b}
\]
Within these tests, $+$ is meant to model disjunction, $\cdot$ models conjunction, and $\overline{\phantom{c}}$ is negation.
Using tests as well as a similarly fixed set of \emph{primitive programs} $\Sigma$, we can then build the set of KAT expressions $\Exp$ as follows:
\[
    \Exp \ni e, f ::= b \in \BExp \mid p \in \Sigma \mid e + f \mid e \cdot f \mid e^*
\]
Here, tests represent assertions: when $b \in \Exp$, $b$ as a program does nothing when $b$ is true, and aborts otherwise.
The operator $+$ represents non-deterministic composition: $e + f$ non-deterministically runs $e$ or $f$; in the same vein, ${}^*$ represents non-deterministic loops: $e^*$ runs $e$ some non-deterministic (possibly zero) number of times.
Finally, $\cdot$ is sequential composition, i.e., $e \cdot f$ first runs $e$, and then $f$.

This syntax is close to that of regular expressions, and of course this is no accident, as Kleene Algebra proper was developed to study their equational theory~\cite{conway-1971}.
We will treat KAT expressions using the precedence rules of regular expressions: ${}^*$ binds more tightly than $\cdot$, which takes priority over $+$.

\subsection{Encoding programs}
One way to think of KAT expressions is as regular expressions instrumented with assertions, describing the ``language'' of traces representing possible ways of executing a program --- we will make this more precise momentarily.
For now though, we should make clear that in spite of this non-determinism, traditional program compositions can still be modelled in KAT~\cite{kozen-1996}.
\begin{itemize}
    \item
    Branches like \texttt{if $b$ then $e$ else $f$} can be modelled with the expression $b \cdot e + \overline{b} \cdot f$.
    Intuitively, this program non-deterministically chooses between asserting that $b$ is true, and then executing $e$, or asserting that $b$ is false, and then executing $f$.
    The branch that matches the truth value of $b$ ``survives''.

    \item
    Loops like \texttt{while $b$ do $e$} can be modelled with the expression ${(b \cdot e)}^* \cdot \overline{b}$.
    The intuition here is that $b \cdot e$ is executed some non-deterministic number of times, with each iteration asserting that the condition $b$ is true before running $e$; finally, when the loop is done, the expression asserts that $b$ is false.
\end{itemize}

With these encodings in mind, the program in \Cref{algorithm:triple-loop} can be written as a KAT expression: $e \cdot (b \cdot f)^* \cdot \overline{b} \cdot (c \cdot g \cdot (b \cdot f)^* \cdot \overline{b})^* \cdot \overline{c}$. 
Here, $e$ stands in for the two lines of code before the first loop, $f$ is the body of the first (and third) loop, and $g$ represents the first three lines of the second loop; $b$ is the condition of the first and third loops, while $c$ is the condition of the second loop.
Similarly, the program in \Cref{algorithm:single-loop} can be encoded as $e \cdot ((b + c) \cdot (b \cdot f + \overline{b} \cdot g))^* \cdot \overline{b+c}$. 

\subsection{Semantics}

To further clarify what we mean when we say that two programs are equivalent regardless of how primitive statements interpreted, we first define semantics that \emph{does} involve the meaning of the statements; next we abstract from this interpretation.
Because KAT programs model non-deterministic programs, it seems most reasonable to model their semantics in terms of relations, connecting initial states of the program to the states that can be reached by running it~\cite{kozen-smith-1996}.

Given a set $S$ of states, suppose we had an interpretation $\sigma$ that assigns to each $p \in \Sigma$ a relation $\sigma(p)$ on $S$ representing the action of $p$.
Furthermore, suppose that we had for each $t \in T$ a predicate $\tau(t) \subseteq T$ describing the states where $t$ is true.
We can then derive a relation $\sem{e}_I$ to model the behavior of the program $e$, provided its primitives are interpreted according to $I = (\sigma,\tau)$:~\cite{kozen-smith-1996}
\begin{mathpar}
\sem{0}_I = \emptyset
\and
\sem{1}_I = \id_S
\and
\sem{t}_I = \{ (s, s) \mid s \in \tau(t) \}
\and
\sem{\overline{b}}_I = \id_S \setminus \sem{b}_I
\and
\sem{p}_I = \sigma(p)
\and
\sem{e + f}_I = \sem{e}_I \cup \sem{f}_I
\and
\sem{e \cdot f}_I = \sem{e}_I \circ \sem{f}_I
\and
\sem{e^*}_I = \sem{e}_I^*
\end{mathpar}
Here, we write $\id_S$ for the identity relation on $S$, and $R^*$ for the reflexive-transitive closure of a relation $R \subseteq S \times S$ --- overloading the Kleene star operator.
The operator $+$ (resp. $\cdot$) is interpreted the same way, whether it is applied to programs or tests.
It is not too hard to prove that $\sem{b}_I$ is always a subset of $\id_S$, for any $b \in \BExp$, and in fact, for $b, c \in \Exp$ we have $\sem{b \cdot c}_I = \sem{b}_I \cap \sem{c}_I$.
At this point, the reader may want to verify that our encoding of traditional flow control aligns with this semantics --- for instance, that $\sem{t \cdot p_1 + \overline{t} \cdot p_2}_I$ is exactly
\[
    \{ (s, s') \mid s \in \tau(t) \wedge (s, s') \in \sigma(p_1) \}
        \cup \{ (s, s') \mid s \not\in \tau(t) \wedge (s, s') \in \sigma(p_2) \}
\]

With all of this in place, our claim from the introduction can finally be made formal: for all interpretations $I$, it holds that the two encodings (presented above) of the programs in \Cref{figure:algorithms} are mapped to the same relation by $\sem{-}_I$.
More broadly, we are interested in all pairs of programs related similarly for all $I$ --- that is to say, in the \emph{equational theory} induced by this semantics.

When further advancing our understanding, quantifying over all interpretations can become cumbersome.
This is where the \emph{language semantics} of KAT comes in~\cite{kozen-smith-1996}.
In a nutshell, the language semantics of a KAT expression gives an accounting of the possible ways the program \emph{could} be executed, without making any assumptions about the interpretations of the primitive symbols.
Among other things, this means that paths of execution that simply cannot happen --- i.e., dead code --- do not contribute to the language semantics.

Executions of a program are recorded in so-called \emph{guarded strings}~\cite{kozen-smith-1996}, which are plain words that alternate between actions and \emph{atoms}.
The latter record the truth value of each test at that point in the program.
More formally, the set of atoms $\At$ is $2^T$, and $\alpha \in \At$ is the atom signifying that all tests in $\alpha$ are true, whereas the primitive tests outside $\alpha$ are false.
A guarded string is a word of the form $\alpha_0 p_0 \alpha_1 p_1 \cdots \alpha_n$, i.e., an element of the regular language $\At \cdot (\Sigma \cdot \At)^*$. 
This guarded string records that the memory of the program was initially described by $\alpha_0$, and after executing $p_0$ the description of the memory was reflected by $\alpha_1$, and so on, until the program finally terminated with memory described by $\alpha_n$.

A set of guarded strings is called a \emph{guarded language}.
To give a semantics of KAT in terms of such languages, we need some formal tools.
Given two guarded languages $L_1$ and $L_2$, their \emph{fusion product} $L_1 \diamond L_2$ is the guarded language $\{ w\alpha{}x \mid w\alpha \in L_1 \wedge \alpha{}x \in L_2 \}$.
If we then write $L^{(0)}$ for $\At$ and $L^{(n+1)}$ for $L^{(n)} \diamond L$, with $L^{(*)} = L^{(0)} \cup L^{(1)} \cup \cdots$, then the guarded language semantics $\sem{e}$ of a KAT expression $e$ can be defined inductively, as follows:~\cite{kozen-smith-1996}
\begin{mathpar}
    \sem{0} = \emptyset
    \and
    \sem{1} = \At
    \and
    \sem{t} = \{ \alpha \in \At \mid t \in \alpha \}
    \and
    \sem{\overline{b}} = \At \setminus \sem{b}
    \and
    \sem{p} = \At \cdot \{ p \} \cdot \At
    \and
    \sem{e+f} = \sem{e} \cup \sem{f}
    \and
    \sem{e \cdot f} = \sem{e} \diamond \sem{f}
    \and
    \sem{e^*} = \sem{e}^{(*)}
\end{mathpar}

The guarded language semantics of KAT is connected to its relational semantics in the intended way: for all $e, f \in \Exp$, we have that $\sem{e} = \sem{f}$ if and only if for all interpretations $I$, we have $\sem{e}_I = \sem{f}_I$~\cite{kozen-smith-1996}.
Importantly, this means that to decide whether $e$ and $f$ coincide relationally for all interpretations $I$ is equivalent to deciding whether their guarded language semantics is the same.

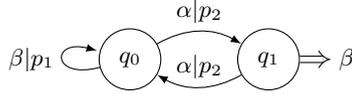
\begin{figure}[t]
    \centering
    \begin{tikzpicture}[>={latex}]
        \node[state] (q0) {$q_0$};
        \node[state,right=of q0] (q1) {$q_1$};
        \draw[-latex] (q0) edge[bend left] node[above] {$\alpha|p_2$} (q1);
        \draw[-latex] (q1) edge[bend left] node[above] {$\alpha|p_2$} (q0);
        \draw[-latex] (q0) edge[loop left] node[left] {$\beta|p_1$} (q0);
        \node[right=4mm of q1] (q1out) {$\beta$};
        \draw (q1) edge[double,double distance=0.7mm,-implies] (q1out);
    \end{tikzpicture}
    \caption{An example automaton on guarded strings}\label{figure:example-automaton}
\end{figure}

Because regular languages correspond precisely to automata, and language equivalence of automata is decidable, this suggests a strategy to decide equivalence in KAT:\@ convert expressions to some kind of automata, and check equivalence there.
The relevant kind of automata are called \emph{automata on guarded strings}; these label transitions with an atom and an action~\cite{kozen-2003,chen-pucella-2004} (see \Cref{figure:example-automaton}).
Moreover, acceptance is indicated for each combination of state and atom; double arrows pointing to an atoms signify acceptance of that atom.
The language of a state is then defined by the guarded strings that are read along transitions to some state, and terminated by an atom accepted at that state.

The conversion from KAT expressions to automata is beyond the scope of this work --- suffice it to say that it is not too different from the well-known conversion of regular expressions to finite automata~\cite{watson-1993}; the procedure to decide language equivalence of (states in) automata on guarded strings is also similar.
We only remark here that checking language equivalence in KAT is \textsc{pspace}-complete~\cite{cohen-kozen-smith-1996}, although a practically efficient algorithm exists~\cite{pous-2015}.

\subsection{Axiomatization}

Besides using the decision procedure sketched above to find out whether two KAT terms are equivalent, we can also reason about them equationally.
The relevant axioms (in \Cref{figure:kat-axioms}) combine Boolean Algebra~\cite{birkhoff-bartee-1970} and Kleene Algebra~\cite{kozen-1994}.

\begin{figure}[t]
    \begin{mathpar}
    b + \overline{b} \equiv 1
    \and
    b \cdot \overline{b} \equiv 0
    \and
    b \cdot b \equiv b
    \and
    b \cdot c \equiv c \cdot b
    \and
    b + c \cdot d \equiv (b + c) \cdot (b + d)
    \and
    e + 0 \equiv e
    \and
    e + e \equiv e
    \and
    e + f \equiv f + e
    \and
    e \cdot 1 \equiv e \equiv 1 \cdot e
    \and
    e \cdot 0 \equiv 0 \equiv 0 \cdot e
    \\
    e + (f + g) \equiv (e + f) + g
    \and
    e \cdot (f \cdot g) \equiv (e \cdot f) \cdot g
    \\
    e \cdot (f + g) \equiv e \cdot f + e \cdot g
    \and
    (e + f) \cdot g \equiv e \cdot g + f \cdot g
    \\
    1 + e \cdot e^* \equiv e^* \equiv 1 + e^* \cdot e
    \and
    e + f \cdot g \leqq g \implies f^* \cdot e \leqq g
    \and
    e + f \cdot g \leqq f \implies e \cdot g^* \leqq f
    \end{mathpar}
    \caption{%
        The axioms of KAT\@.
        Here, $\equiv$ is the smallest congruence on KAT that satisfies the laws above for all $e, f, g \in \Exp$ and $b, c, d \in \BExp$.
        We also write $e \leqq f$ as a shorthand for $e + f \equiv f$; this makes $\leqq$ a partial order (up to $\equiv$).
    }\label{figure:kat-axioms}
\end{figure}

It is not too difficult to prove soundness of these axioms: when $e \equiv f$, it also holds that $\sem{e} = \sem{f}$, by induction on $\equiv$.
In contrast, \emph{completeness} --- i.e., when $\sem{e} = \sem{f}$, we have that $e \equiv f$ --- is far from trivial~\cite{kozen-smith-1996}.
Known proofs all seem to hinge on a version of Kleene's theorem for KAT.%
\footnote{
    More accurately, the usual proof of completeness for KAT~\cite{kozen-smith-1996} relies on a reduction to completeness of Kleene Algebra~\cite{kozen-1994}.
    The latter typically follows a tactic very close to the one outlined here~\cite{kozen-1994,kozen-2001,jacobs-2006}.
    Importantly, the two-way correspondence between expressions and automata is crucial in alternative proofs, too~\cite{kappe-2024}.
}
An example of such a proof follows.
First, we cast automata as systems of equations in KAT, with a variable for each state.
For example, the automaton in \Cref{figure:example-automaton} corresponds to
\begin{mathpar}
    \alpha \cdot p_2 \cdot x_1 + \beta \cdot p_1 \cdot x_0 \equiv x_0
    \and
    \alpha \cdot p_2 \cdot x_0 + \beta \equiv x_1
\end{mathpar}
where $x_i$ is the variable for state $q_i$, and we used atoms $\gamma$ to denote the test asserting that $t \in T$ is true if and only if $t \in \gamma$.
Such a system of equations admits a \emph{least} solution (w.r.t. $\leqq$)~\cite{conway-1971,kozen-1994}; this can be seen as an algebraic take on the construction to convert an automaton to an expression~\cite{mcnaughton-yamada-1960}.
Moreover, we have a \emph{round trip theorem}: the least solution to the variable for the initial state in the system obtained from the automaton for $e$ is always equivalent to $e$~\cite{kozen-1994}.

Now, given an expression $e$ (resp. $f$), we can come up with an automaton on guarded strings for $e$ (resp. $f$), whose language is $\sem{e}$ (resp. $\sem{f}$).
Since $\sem{e} = \sem{f}$, the automata for $e$ and $f$ have the same language.
With some effort, we can then prove that the systems of equations corresponding to equivalent automata must have equivalent least solutions~\cite{kozen-1994,jacobs-2006}.
This allows us to conclude that $e$ and $f$ are equivalent, because the least solutions to their automata are, too.

\section{Determinism}\label{section:gkat}

With the theory discussed so far, it is possible to encode the two programs from \Cref{figure:algorithms}, and then either mechanically check that they are equivalent, or prove that this is the case using the laws from \Cref{figure:kat-axioms}.
But why should we need to use non-determinism to encode control flow that is essentially deterministic?
To explore this question further, we shift gears to talk about GKAT~\cite{kozen-tseng-2008,smolka-etal-2020}, the fragment of KAT that can be built using the embeddings of \texttt{if $b$ then $e$ else $f$} and \texttt{while $b$ to $e$}.
Formally, the syntax of GKAT is generated as follows:
\[
    \GExp \ni e, f ::= b \in \BExp \mid p \in \Sigma \mid e +_b f \mid e \cdot f \mid e^{(b)}
\]
We use $e +_b f$ as a shorthand for \texttt{if $b$ then $e$ else $f$}, and $e^{(b)}$ to denote \texttt{while $b$ do $e$}.
Also, ${}^{(b)}$ takes precedence over $\cdot$, which has a higher priority than $+_b$.

The semantics of GKAT can be defined by embedding of GKAT into KAT, sending $e +_b f$ to $b \cdot e + \overline{b} \cdot f$, and $e^{(b)}$ to $(b \cdot e)^* \cdot \overline{b}$~\cite{kozen-1996}. 
Going forward, when $e$ is a GKAT expression we will simply write $\sem{e}$ (resp.\ $\sem{e}_I$) for the guarded language obtained by viewing $e$ as a KAT expression and taking its guarded language (resp.\ relational semantics w.r.t.\ $I$).
Alternatively, we can interpret the primitive actions in $\Sigma$ as functional relations (i.e., partial functions)~\cite{tencate-kappe-2025}.
If $I = (\sigma, \tau)$ with $\sigma(p)$ a functional relation for each $p \in \Sigma$, then $\sem{e}_I$ is also a partial function for each $e \in \GExp$.\footnote{%
    We can also directly define a partial function semantics of GKAT expressions, e.g., by sending $e \cdot f$ to $\sem{f}_I \circ \sem{e}_I$~\cite{tencate-kappe-2025}.
    The only slightly tricky (but still doable) part is the semantics of loops, which require directed-completeness of partial function inclusion.
}
The connection between the language semantics and relational semantics specializes to partial functions for GKAT:\@ for all $e, f \in \GExp$, we have that $\sem{e} = \sem{f}$ if and only if $\sem{e}_I = \sem{f}_I$ for each functional interpretation $I$.

By treating these constructs as first-class citizens, the automata obtained satisfy a special property: they are \emph{deterministic}~\cite{kozen-tseng-2008,smolka-etal-2020}, in the sense that for every atom there is exactly one possibility between accepting, not accepting, or transitioning to a state.
Furthermore, when an atom causes a transition to another state, it does so with only one action, and to only one state.
The automaton drawn in \Cref{figure:example-automaton} is an example of a deterministic automaton.
Note that an automaton on guarded strings can be classically deterministic (in that there is at most one transition exiting each state with a given atom and action), while being non-deterministic in our sense --- for instance, by having a transition labeled $\alpha|p_1$ to one state, and one labeled $\alpha|p_2$ to another (or the same) state.

This property of automata obtained for GKAT expressions gives rise to a specialized conversion from expressions to automata, in the style of Thompson~\cite{thompson-1968}.
Unlike the general conversion from KAT expressions to automata on guarded strings, this does \emph{not} involve a determinization step, and hence avoids the exponential blowup that may occur.
Indeed, the automata produced have a number of states that is \emph{linear} in the size of the expression~\cite{smolka-etal-2020}.
We can combine this with an efficient algorithm to check language equivalence of automata~\cite{hopcroft-karp-1971} to obtain a decision procedure whose complexity is \emph{nearly linear} (thanks to the union-find data structure~\cite{tarjan-1975}) in the size of the expressions involved~\cite{smolka-etal-2020}.

\subsection{Axiomatization}

\begin{figure}[t]
    \begin{mathpar}
    e +_b e \equiv e
    \and
    e +_b f \equiv f +_{\overline{b}} e
    \and
    (e +_b f) +_c \equiv e +_{b \cdot c} (f +_c g)
    \and
    e +_b f \equiv b \cdot e +_b f
    \\
    e \cdot g +_b f \cdot g \equiv (e +_b f) \cdot g
    \and
    (e \cdot f) \cdot g \equiv e \cdot (f \cdot g)
    \and
    0 \cdot e \equiv 0 \equiv e \cdot 0
    \and
    1 \cdot e \equiv e \equiv e \cdot 1
    \and
    e^{(b)} \equiv e \cdot e^{(b)} +_b 1
    \and
    (c \cdot e)^{(b)} \equiv (e +_c 1)^{(b)} 
    \end{mathpar}
    \caption{%
        Some axioms of GKAT~\cite{smolka-etal-2020}.
        Here, $\equiv$ is the smallest congruence on $\GExp$ generated by the laws above, for all $e, f, g \in \GExp$ and $b, c \in \BExp$.
    }\label{figure:gkat-axioms}
\end{figure}

Because GKAT specializes KAT, a natural question is whether the same type of results can be achieved.
Previously, we already indicated that an efficient decision procedure exists.
Moreover, we can also translate GKAT expressions to KAT expressions, and reason about them there --- but can we prove these equivalences \emph{within GKAT}?
To this end, we need laws expressed in the language of GKAT itself; some candidates appear in \Cref{figure:gkat-axioms}~\cite{smolka-etal-2020}.
For instance, $e \cdot g +_b f \cdot g \equiv (e +_b f) \cdot g$ says that actions common to the end of a conditional can be factored out.
Each of these laws can be shown to be sound w.r.t.\ the semantics~\cite{smolka-etal-2020}.

One thing missing from these laws is a description of loops as a least fixed point, like how KAT describes $e^* \cdot f$ as the least fixed point (w.r.t. $\leqq$) of $x \mapsto e \cdot x + f$.
The only problem is that, unlike KAT, GKAT does not have a native partial order --- for the simple reason that it lacks the $+$ operator from KAT\@.

The first way to resolve this issue is to axiomatize loops as \emph{unique} fixed points.
To a first approximation, it may seem reasonable to stipulate that, if $g \equiv e \cdot g +_b f$, we should be able to conclude that $g \equiv e^{(b)} f$ --- after all, $g$ seems to be a program that, when $b$ is true, executes $e$ and then restarts, and otherwise executes $f$.
Unfortunately, this rule would be unsound; take for instance $e = f = g = b = 1$: in this case, the premise is true by the laws in \Cref{figure:gkat-axioms}, but the consequence $1 \equiv 1^{(1)} \cdot 1$ cannot be accepted.
Intuitively, this is because $1$ succeeds immediately, while $1^{(1)} \cdot 1$ describes a (non-productive) infinite loop; formally, the relational semantics maps the former to the identity function, and the latter to the empty partial function.
More broadly, this issue occurs when the loop body $e$ may be non-productive.
To get around this problem, we can take inspiration from the approach championed by Salomaa~\cite{salomaa-1966}, and define a side-condition to our rule to stipulate that the loop body must be productive.
This can be done by defining a function $E: \GExp \to \BExp$ that sends each GKAT program $e$ to a test $E(e)$, describing the circumstances under which $e$ can be ``skipped'' --- i.e., executed without performing any action.
Formally, $E$ is defined as follows.
\begin{mathpar}
    E(b) = b
    \and
    E(p) = 0
    \and
    E(e +_b f) = b \cdot E(e) + \overline{b} \cdot E(f)
    \and
    E(e \cdot f) = E(e) \cdot E(f)
    \and
    E(e^{(b)}) = \overline{b}
\end{mathpar}
We can now state our \emph{unique fixed point} rule as follows: if $g \equiv e \cdot g +_b f$ and $E(e) \equiv 0$, then $g \equiv e^{(b)} \cdot f$.
This prevents the earlier pathological case~\cite{smolka-etal-2020}.

An alternative approach to the lack of order is to not axiomatize equivalence, but rather inclusion of programs --- given $e$ and $f$, we could look at when $\sem{e} \subseteq \sem{f}$.
To achieve this, we can simply think of the axioms in \Cref{figure:gkat-axioms} as syntactic sugar for two inclusions of this form, i.e., $e +_b e \equiv e$ stands in for $e +_b e \leqq e$ \emph{and} $e \leqq e +_b e$.
With that in mind, we can define a least fixed point rule to state that, if $e \cdot g +_b f \leqq g$, then $e^{(b)} \cdot f \leqq g$; this rule is sound w.r.t.\ the semantics.

Although both approaches produce a sound reasoning system, it is an open problem whether they are \emph{complete}, i.e., whether they can be used to prove any valid equivalence or containment.
The root of the problem is that the systems of equations induced by deterministic automata do not always admit a GKAT expression as a solution~\cite{kozen-tseng-2008} (we will explore this issue momentarily).
As such, the strategies to prove completeness for KAT, which all hinge on being able to solve systems of equations, cannot be cleanly applied in the setting of GKAT\@.

One way to circumvent this issue and obtain a complete axiomatization is to strengthen the unique fixed point axiom.
This axiom states that an equation with one unknown of the form $x \equiv e \cdot x +_b f$ has exactly one solution up to equivalence, namely $e^{(b)}$, provided that $E(e) \equiv 0$.
While systems with more than one unknown may not admit any solution, it is sound to say that finite systems of equations with more than one variable (in a similar format), admit \emph{at most} one solution, provided a similar side condition is satisfied~\cite{smolka-etal-2020}.
This leads to an infinite number of rules, one for each number of variables; including this infinitary axiom scheme is sufficient to axiomatize equivalence of GKAT~\cite{smolka-etal-2020}.

\subsection{Expressivity}

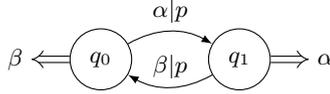
\begin{figure}[t]
    \centering
    \begin{tikzpicture}[>={latex}]
        \node[state] (q0) {$q_0$};
        \node[state,right=of q0] (q1) {$q_1$};
        \node[left=5mm of q0] (q0acc) {$\beta$};
        \node[right=5mm of q1] (q1acc) {$\alpha$};
        \draw[-latex] (q0) edge[bend left] node[above] {$\alpha|p$} (q1);
        \draw[-latex] (q1) edge[bend left] node[above] {$\beta|p$} (q0);
        \draw (q0) edge[double,double distance=0.7mm,-implies] (q0acc);
        \draw (q1) edge[double,double distance=0.7mm,-implies] (q1acc);
    \end{tikzpicture}
    \caption{A deterministic automaton on guarded strings without a solution~\cite{schmid-etal-2021}.}\label{figure:example-bad-automaton}
\end{figure}

A moment ago, we noted that not all deterministic automata on guarded strings give rise to a system of equations solvable within GKAT~\cite{kozen-tseng-2008,schmid-etal-2021}.
An example of such an automaton is depicted in \Cref{figure:example-bad-automaton}.
Intuitively, the problem there is that there is no condition that terminates the loop: on the left, the atom $\alpha$ resumes the loop while $\beta$ terminates execution, whereas on the right this is reversed~\cite{schmid-etal-2021}.

One might wonder whether the behavior exemplified by this automaton is the \emph{only} deterministic behavior not expressed by GKAT --- in other words, can we perhaps extend our language with an operation $\mathsf{twostate}(b, c, e)$ corresponding to this automaton?
Unfortunately, extending our language like this gets us only so far: it can be shown that there is an infinite hierarchy of automata like this one, each with behavior inexpressible in terms of the ones that came before~\cite{tencate-kappe-2025}.
Thus, a finite extension of the language of GKAT will not allow us to express all deterministic behaviors.
The best we can hope for is a characterization of the class of automata on guarded strings that can be solved within GKAT\@.

The strongest result that we have to relate GKAT expressions to deterministic automata on guarded strings goes as follows.
First, given a deterministic automaton on guarded strings $A$ that is \emph{well-nested} (in a sense not elaborated upon here), we can find a solution $e$~\cite{smolka-etal-2020}.
Conversely, for each GKAT expression $e$, we can derive a well-nested deterministic automaton on guarded strings $A$~\cite{smolka-etal-2020,kozen-tseng-2008}.
While this is a two-way correspondence between a fragment of deterministic automata on guarded strings, it is somewhat unsatisfactory because well-nestedness is only a sufficient condition for the existence of a solution --- we know of automata that do admit a solution, but which are not well-nested~\cite{schmid-etal-2021}.

A proper characterization of solvable automata would go a long way towards proving a completeness theorem for GKAT, simply because unique (or least) solutions to automata are such a powerful tool.
At the time of writing, a characterization along these lines remains elusive, and so our labor continues.

\smallskip
The problem of characterizing solvable automata is also related to the age-old discussion of which programs can (not) be expressed without \texttt{goto}~\cite{dijkstra-1968,dijkstra-1987}.
We could therefore think about extending GKAT with \emph{non-local control flow}, i.e., primitives that affect control flow dependent on their context.
The automaton in \Cref{figure:example-bad-automaton} could be captured by a program like \texttt{while $\alpha$ do \{ $p$; if $\alpha$ then break; $p$ \}}.
In particular, if we include a \texttt{goto} statement in our syntax, then \emph{all} deterministic GKAT automata can be converted to expressions.
While such an extension maintains fairly favorable decidability properties~\cite{zhang-etal-2025}, it is unclear how the context-dependent semantics of the added control flow statements can be incorporated in the axiomatization.
More advanced reasoning techniques, beyond the traditional algebraic ones, may be necessary.

\subsection{Bisimilarity}

\begin{figure}[t]
    \centering
    \begin{mathpar}
    \inferrule{%
        \alpha \leqq E(e)
    }{%
        e \Downarrow \alpha
    }
    \and
    \inferrule{~}{%
        p \xrightarrow{\alpha|p} 1
    }
    \and
    \inferrule{%
        e \xrightarrow{\alpha|p} e'
        \and
        \alpha \leqq b
    }{%
        e +_b f \xrightarrow{\alpha|p} e'
    }
    \and
    \inferrule{%
        f \xrightarrow{\alpha|p} f'
        \and
        \alpha \leqq \overline{b}
    }{%
        e +_b f \xrightarrow{\alpha|p} f'
    }
    \and
    \inferrule{%
        e \xrightarrow{\alpha|p} e'
    }{%
        e \cdot f \xrightarrow{\alpha|p} e' \cdot f
    }
    \and
    \inferrule{%
        e \Downarrow \alpha
        \and
        f \xrightarrow{\alpha|p} f'
    }{%
        e \cdot f \xrightarrow{\alpha|p} f'
    }
    \and
    \inferrule{%
        e \xrightarrow{\alpha|p} e'
        \and
        \alpha \leqq b
    }{%
        e^{(b)} \xrightarrow{\alpha|p} e' \cdot e^{(b)}
    }
    \end{mathpar}
    \caption{%
        Operational semantics of GKAT terms~\cite{schmid-etal-2021}. Here, $\alpha \in \At$, $e, f \in \GExp$, and $b \in \BExp$. We use $\leqq$ to denote containment (implication) in Boolean algebra.}\label{figure:transition-rules}
\end{figure}

One reasonable objection to the axioms and semantics of GKAT is that it equates programs that fail immediately (e.g., $0$) to programs that inevitably fail later (e.g., $e \cdot 0$).
This becomes particularly apparent when we realize that the axioms of GKAT allow us to prove that $e^{(b)} \equiv e^{(b)} \cdot \overline{b}$ --- after all, a loop condition should be false when the loop has ended --- and thus $e^{(1)} \equiv e^{(1)} \cdot 0 \equiv 0$.
In other words, infinite loops are equivalent to programs that fail immediately~\cite{schmid-etal-2021}.

To address this problem, we can instead choose to study the equivalence on GKAT terms induced by \emph{bisimilarity} in the transition system generated by the rules in \Cref{figure:transition-rules}.
More precisely, $e$ and $f$ are bisimilar if there exists a bisimulation that relates them, i.e., a relation $R$ on $\GExp$ such that if $e' \mathrel{R} f'$, then $e' \Downarrow \alpha$ implies $f' \Downarrow \alpha$, and if $e' \xrightarrow{\alpha|p} e''$, then $f' \xrightarrow{\alpha|p} f''$ with $e'' \mathrel{R} f''$.\footnote{%
    Because of determinism, the converse requirement (``back condition'') usually included for bisimilarity turns out not to be necessary~\cite{smolka-etal-2020}.
}

The results that we do have regarding axiomatization of GKAT --- i.e., that a generalized unique fixed point rule suffices to obtain completeness --- can be recovered for this setting.
The only requirement is that we drop the right-annihilation axiom $e \cdot 0 \equiv 0$, because it is incompatible with bisimilarity ($p \cdot 0$ is not bisimilar to $0$)~\cite{schmid-etal-2021}.
Indeed, from this result we can even recover the original claim about the axiomatization of the language semantics~\cite{schmid-etal-2021}.
This suggests that bisimilarity might be more interesting as a primary equivalence for GKAT, which can then be coarsened when language equivalence is more desirable.

%

\section{Connections to process algebra}\label{section:pa}

\begin{figure}[t]
    \centering
    \begin{mathpar}
        \inferrule{~}{%
            1 \downarrow
        }
        \and
        \inferrule{
            e \downarrow
        }{%
            e + f \downarrow
        }
        \and
        \inferrule{
            f \downarrow
        }{%
            e + f \downarrow
        }
        \and
        \inferrule{
            e \downarrow
            \and
            f \downarrow
        }{%
            e \cdot f \downarrow
        }
        \and
        \inferrule{~}{%
            e^* \downarrow
        }
        \and
        \inferrule{
            p \in \Sigma
        }{%
            p \xrightarrow{p} 1
        }
        \and
        \inferrule{%
            e \xrightarrow{p} e'
        }{%
            e + f \xrightarrow{p} e'
        }
        \and
        \inferrule{%
            f \xrightarrow{p} f'
        }{%
            e + f \xrightarrow{p} f'
        }
        \and
        \inferrule{%
            e \xrightarrow{p} e'
        }{%
            e \cdot f \xrightarrow{p} e' \cdot f
        }
        \and
        \inferrule{%
            e \downarrow
            \and
            f \xrightarrow{p} f'
        }{%
            e \cdot f \xrightarrow{p} f'
        }
        \and
        \inferrule{%
            e \xrightarrow{p} e'
        }{%
            e^* \xrightarrow{p} e' \cdot e^*
        }
    \end{mathpar}
    \caption{%
        Transition rules for regular expressions.
        All of these are quantified over regular expressions $e$ and $f$, and letters $p \in \Sigma$.
    }\label{figure:regular-transition-rules}
\end{figure}

We finish with some exposition on a potential connection to process algebra, and how recent results there can be leveraged to partially answer our questions.

Language equivalence of regular expressions has a well-known axiomatization in the form of Kleene Algebra~\cite{kozen-1994}.
However, \emph{bisimilarity} of regular expressions is a different beast altogether, and one that has eluded characterization for quite some time.
This problem, first posed by Milner~\cite{milner-1984}, revolves around the transition system on regular expressions induced by the rules in \Cref{figure:regular-transition-rules}.

It is not too hard to see that, if $e$ and $f$ are bisimilar, then they represent the same regular language.
Nevertheless, language equivalence does not imply bisimilarity --- consider for instance the expressions $p_1 \cdot (p_2 + p_3)$ and $p_1 \cdot p_2 + p_1 \cdot p_3$.
This example also tells us that the usual distributivity rule for regular expressions does not apply in this setting.
Milner recognized this, and conjectured that dropping distributivity and the right annihilation rule $e \cdot 0 \equiv 0$ would give rise to a complete axiomatization of bisimilarity for regular expressions~\cite{milner-1984}.

However, when adapting Salomaa's proof for completeness of regular expressions, an issue very similar to the one encountered above arises: there exist non-deterministic automata that are not bisimilar to any regular expression~\cite{milner-1984}.
This lack of a proper back-and-forth correspondence between expressions and automata proved to be a roadblock towards proving completeness.

Relatively recent work by Grabmayer and Fokkink has made an important step in resolving Milner's conjecture: if we restrict ourselves to the $1$-free fragment of regular expressions, i.e., where the constant $1$ does not appear and the Kleene star is restricted to be a binary operator (written $e^*f$), then bisimilarity can be axiomatized~\cite{grabmayer-fokkink-2020}.
Indeed, this result hinges on carving out a class of automata that corresponds to regular expressions up to bisimilarity, but that is also stable under bisimulation collapse~\cite{grabmayer-fokkink-2020} (and in fact, under homomorphic images~\cite{schmid-rot-silva-2021}).
This property is key to axiomatizing bisimilarity in this fragment.

The theme of characterizing a certain class of automata suggests that perhaps these results about $1$-free regular expressions can be adapted to the setting of GKAT, and indeed: a suitably restricted fragment of GKAT, called \emph{skip-free GKAT}, can be axiomatized using the same techniques~\cite{kappe-schmid-silva-2023}.
Crucially, the result that the class of automata under study is closed under bisimulation collapse, which is extremely tricky to prove, can be reused as-is.
Moreover, the techniques can be generalized using a coalgebraic approach to obtain an abstract completeness result that characterizes equivalence of $1$-free regular expressions, skip-free GKAT, and even the skip-free fragment of GKAT's probabilistic extension~\cite{kappe-schmid-2025}.

Finally, a full proof of Milner's conjecture was announced by Grabmayer~\cite{grabmayer-2022}.
While the arguments in that work are highly intricate, and more work is necessary to clarify the details in the accompanying appendices, we remain hopeful that the mathematical essence of this work has the potential for abstraction along the same way as the preceding work on $1$-free regular expressions.

\section{New horizons}\label{section:horizons}

Besides the two central questions on axiomatization and equivalence, several other worthwhile avenues of research exist.
We highlight a few of these now.

First, we can flip the question about expressivity to ask whether we can restrict KAT to express \emph{all} deterministic automata.
As indicated before, this is not possible by just adding deterministic composition operators that occur within KAT itself~\cite{tencate-kappe-2025}.
Nevertheless, it may be possible to devise new deterministic composition operators that do not have an equivalent in KAT\@.
One candidate for this is the \emph{domain} operator, which turns a program $e$ into a test $D(e)$ that describes the circumstances under which $e$ may (eventually) terminate successfully~\cite{desharnais-moeller-struth-2006,sedlar-2023}.

Intuitively, we can think of $D(e)$ as a program that calls a function whose body is $e$, and returns whether $e$ was able to terminate successfully.
The hope is that this operator could potentially be used to achieve the deterministic behavior that is normally modelled in KAT by non-deterministic compositions that turn out to behave deterministically, because at most one branch survives.
In essence, $D$ might help resolve non-determinism ahead of time, by peeking ahead.

Second, one could try and take another look at the behaviors expressible by GKAT terms from a coalgebraic point of view.
Collectively, these behaviors induce a \emph{covariety} of automata~\cite{schmid-etal-2021}.
The automata in this covariety can be viewed as Eilenberg-Moore coalgebras for a comonad~\cite{goldblatt-2005}.
One may then ask: can we describe the \emph{coequations} that characterize this comonad?
While these notions have well-understood duals (varieties, monads, Eilenberg-Moore algebras), their mathematics leave something to be desired.
In particular, there are different methods for ``writing down'' a coequation~\cite{dahlqvist-schmid-2021}, and it is not clear which of these (if any) would be appropriate to characterize the automata of GKAT expressions.

Third, Kleene algebra can be extended with hypotheses~\cite{hardin-2005,doumane-etal-2019,kozen-mamouras-2014,kappe-etal-2020,pous-wagemaker-2024,pous-rot-wagemaker-2024}.
In this setting, we have one or more hypotheses $H = \{ e_1 \equiv f_1, \dots, e_1 \equiv f_n \}$, and we ask whether $H$, in conjunction with the usual laws of Kleene algebra, is sufficient to prove some equivalence $e \equiv f$.
In general, this problem is not decidable~\cite{kozen-1996,kuznetsov-2023,amorim-zhang-gaboardi-2025}, but for special cases, such as when $f_i = 0$ for all $i$, it is~\cite{cohen-1994}.
Moreover, completeness results can also be obtained, usually via a series of reductions to the base completeness theorem for Kleene algebra~\cite{pous-rot-wagemaker-2024}.
We wonder whether similar results can be obtained for GKAT, by leveraging the existing completeness theorem that relies on the infinitary unique solutions axiom.
Unfortunately, the answer is not obvious --- whereas completeness results about KA with hypotheses typically rely on \emph{saturating} an expression to include all of the behavior deemed equivalent by the hypotheses~\cite{doumane-etal-2019}, such an approach is not immediately possible in GKAT, for lack of the $+$ operator.

Fourth, perhaps the most applied direction to take GKAT lies in analyzing (de)compilers. 
On the one hand, a compiler typically performs a great number of non-trivial program transformations; perhaps the theory of GKAT can aid in verifying some of these --- as was done with KAT~\cite{kozen-patron-2000}.
On the other hand, \emph{decompilers} are tasked with analyzing binary code, with the aim of obtaining a more high-level representation of the intended behavior in a programming language~\cite{cifuentes-1994}.
Part of the challenge there lies in structuring control flow that is essentially represented in the form of an automaton.
New developments in characterizations of solvable automata could help improve decompilers.
Conversely, we believe there is much to be learned from modern decompilers, and that perhaps GKAT could one day serve as the theoretical backbone of a verified decompiler~\cite{zhang-etal-2025}.

Finally, there are interesting developments further afield, such as probabilistic~\cite{rozowski-etal-2023}, relational~\cite{gomes-etal-2025} and weighted~\cite{vankoevering-etal-2025} extensions of GKAT, as well as connections to modal logic~\cite{benevides-etal-2024} and cyclic proof systems~\cite{rooduijn-etal-2024}.

\paragraph{Acknowledgements}
I am grateful to Balder ten Cate, Nate Foster, Justin Hsu, Dexter Kozen, David E. Narváez, Nico Naus, Wojciech Różowski, Todd Schmid, Alexandra Silva, Steffen Smolka, and Cheng Zhang for the wonderful collaborations on this topic, and Todd Schmid in particular for helpful commentary on a first draft of this manuscript.
I would furthermore like to thank Jurriaan Rot and Jana Wagemaker for their thoughtful insights, and Hendrik Jan Hoogeboom for proposing the example of the two in-order tree traversal algorithms.

This work was supported by the Dutch Research Council (NWO) under grant no. VI.Veni.232.286 (ChEOpS).

\bibliographystyle{splncs04}
\bibliography{bibliography}
\end{document}